\documentclass{article}

\usepackage{charter}

\usepackage{amsmath}
\usepackage{amssymb}
\usepackage{amsfonts,amssymb,latexsym}
\usepackage{graphicx}

\usepackage{subfigure}
\usepackage{tikz}
\usetikzlibrary{arrows}
\usepackage{gantt}

\newtheorem{Theo}{Theorem}

\newcommand{\dbf}{\textsf{dbf}}
\newcommand{\hbf}{\textsf{hbf}}

\begin{document}

	\title{Comments on ``Gang EDF Schedulability Analysis''}

	\author{Pascal~Richard\thanks{LIAS-University of Poitiers, pascal.richard@univ-poitiers.fr}\and Jo\"el~Goossens\thanks{ Universit\'e Libre de Bruxelles, Belgium, joel.goossens@ulb.ac.be} \and Shinpei~Kato \thanks{Nagoya University, Japan, shinpei@is.nagoya-u.ac.jp}}

    \date{}
	\maketitle

	\begin{abstract}
		This short report raises a correctness issue in the schedulability test presented in \cite{Kato09}: ``Gang EDF Scheduling of Parallel Task Systems'', 30th IEEE Real-Time Systems Symposium, 2009, pp. 459-468.

	\end{abstract}


	\section{Introduction}

	We raise a correctness issue in the schedulability test  in the paper: ``Gang EDF Scheduling of Parallel Task Systems'', Shinpei Kato and Yutaka Ishikawa, presented at RTSS'09. This paper presents a Gang scheduling algorithm (Gang EDF) and its schedulability test, named  \textbf{[KAT]} hereafter.

	\textbf{[KAT]} considers $n$ preemptive sporadic gang tasks (also known as rigid parallel tasks \cite{dutot2004scheduling}), to be executed upon $m$ identical processors. Each task $\tau_i=(v_i, C_i,D_i,T_i)$, $1 \leq i \leq n$, is characterized by the number $v_i$ of used processors, a worst-case execution time $C_i$ when executed in parallel on $v_i$ processors, a minimum inter-arrival time $T_i$ and a constrained relative deadline $D_i$ (i.e., $D_i \leq T_i$). The utilization of $\tau_i$ is $U_i=C_i/T_i$. Each task generates an infinite sequence of jobs. The execution of a job of $\tau_i$ is represented as a $C_i \times v_i$ rectangle in time $\times$ processor space. Every job must be completed by its deadline.

	Gang EDF applies the well-known EDF policy to the gang scheduling problem: jobs with earlier deadlines are assigned higher priorities. Since tasks use several processors, the earliest deadline rule is extended in Gang EDF to take into account the spacial limitation on the number of available processors by using a first fit based strategy. The reader can refer to~\cite{Kato09} for a complete description of the scheduling algorithm. Next, we only focus on the schedulability test for analyzing gang tasks scheduled by Gang EDF.

	\section{[KAT] Schedulability Test Principles}\label{sec:princ}

	Basically, the test follows the principles of \textbf{[BAR]} test defined in~\cite{Bar07} for tasks using at most one processor. It is based on a \textit{necessary schedulability condition} for a task to miss a deadline. Then, the contrapositive condition yields a \textit{sufficient schedulability condition} for the considered scheduling algorithm.

	\textbf{[KAT]} test considers any legal sequence of job requests on which a deadline is missed by Gang~EDF. Assume that $\tau_k$ is generating the problem job at time $t_a$ that must be completed by its deadline at time $t_d=t_a+D_k$. Let $t_0$ be the latest time instant before or at  $t_a$ at which at least $v_k$ processors are idled and $\Delta_k=t_d-t_0$. A necessary condition for the problem job to miss its deadline is: higher priority tasks are blocking $\tau_k$ for strictly more that $D_k-C_k$ in the interval $[t_a,t_d)$. Since $\tau_k$ requires $v_k$ processors, it is blocked while $m-v_k+1$ processors are busy. This minimum interference necessary for the deadline miss is defined by the \textit{interference rectangle} whose width $w_k$ and height $h_k$ are respectively given by: $w_k = \Delta_k -C_k$ and $h_k = m-v_k+1$.

	Let $I(\tau_i,\Delta_k)$ be the worst-case interference against the problem job over $[t_0,t_d)$, meaning that it blocks the problem job over $[t_a,t_d)$ and is executed over $[t_0,t_a)$. In~\cite{Kato09}, it is claimed that: \textit{\textbf{If the problem job misses its deadline, then the total amount of work that interferes over $[t_0,t_d)$ necessarily exceeds the interference rectangle:}}

	\begin{eqnarray}
		\sum_{\tau_i \in \tau} I(\tau_i,\Delta_k) > w_k\times h_k \label{eq:pb}
	\end{eqnarray}

	It is important to understand exactly what Equation (\ref{eq:pb}) means: \textbf{if a task misses a deadline, then the condition defined in (\ref{eq:pb}) is satisfied}. Thus, it represents necessary conditions for task $\tau_k$ to miss a deadline $A_k=\Delta_k-D_k$ units after an instant at which at least $v_k$ processors are idled. It is also important to notice that the necessary condition only exploits the area of the interference rectangle (i.e., $w_k \times h_k$ ) for defining a bound on the processor demand for task $\tau_k$. Finally, it is also worth noticing that this necessary condition is not formally proved in~\cite{Kato09}.

	As in \textbf{[BAR]}, the contrapositive of the previous necessary condition allows to define the following sufficient schedulability condition in \textbf{[KAT]}: if the contrapositive of Equation (\ref{eq:pb}) is satisfied, then deadlines are met.

	The interference must take into account carry-in jobs in the interference  rectangle who arrive before $t_0$ and have not completed execution by $t_0$. The \textbf{[KAT]} test distinguishes the interference coming from tasks without or with a carry-in job for bounding the overall interference $I(\tau_i,\Delta_k)$, respectively denoted $I_1(\tau_i,\Delta_k)$ and $I_{\operatorname{carry-in}}$. The schedulability test \textbf{[KAT]} checks every task using the contrapositive of the necessary condition (\ref{eq:pb}) that yields a sufficient schedulability condition in Theorem~\ref{th:test}.

	\begin{Theo}\label{th:test}~\cite{Kato09}
		It is guaranteed that a task system $\tau$ is successfully scheduled by Gang~EDF upon $m$ processors, if the following condition is satisfied for all tasks $\tau_k \in \tau$ and all $\Delta_k \geq D_k$:
		\begin{equation} \label{eq:test}
			\sum_{\tau_i \in \tau} I_1(\tau_i,\Delta_k) + I_{\operatorname{carry-in}} \leq  w_k\times h_k
		\end{equation}
	\end{Theo}

	Several bounds have been proposed in~\cite{Kato09} to evaluate the accumulated interference $\sum_{\tau_i \in \tau} I_1(\tau_i,\Delta_k)$. We limit ourselves to use the first proposed bound:
	\begin{eqnarray*}
		I_1(\tau_i)&=& \min(\hbf(\tau_i,\Delta_k),w_k)\times \min(v_i,h_k) \qquad \qquad \hbox{if $i\neq k$}\\
		I_1(\tau_i)&=& \min(\hbf(\tau_i,\Delta_k)-C_k,A_k)\times \min(v_i,h_k) \qquad \hbox{if $i=k$}
	\end{eqnarray*}
	where $A_k=\Delta_k-D_k$ defines the maximum contribution of $\tau_k$ in the feasibility interval $[t_0,t_d)$ and $\hbf(\tau_i,L)=\max \left( 0, \left\lfloor \frac{L-D_i}{T_i} \right\rfloor  +1 \right)\times C_i $ is the horizontal-demand bound function \footnote{The horizon-demand bound function computes the upper bound of the \emph{time length demand} of $\tau_i$ over any time interval of length $L$~\cite{Kato09}: $\hbf(\tau_i,L)=\dbf(\tau_i,L)\times \frac{1}{v_i}$.}.

	Similarly, the contribution of task $\tau_i$ with a carry-in job to the interference rectangle can be bounded by~\cite{Kato09}:
		\begin{eqnarray*}
			I_2(\tau_i)&=& \min(\hbf'(\tau_i,\Delta_k),w_k)\times \min(v_i,h_k) \qquad \qquad \hbox{if $i\neq k$}\\
			I_2(\tau_i)&=& \min(\hbf'(\tau_i,\Delta_k)-C_k,A_k)\times \min(v_i,h_k) \qquad \hbox{if $i=k$}
		\end{eqnarray*}
	where $\hbf'(\tau_i,L)= \left\lfloor \frac{L}{T_i} \right\rfloor \times C_i + min(C_i,L \mod T_i)$.  $I_{\operatorname{diff}}=I_2(\tau_i)-I_1(\tau_i)$ is the contribution of $\tau_i$ by its carry-in job to the interference rectangle (defined in~\cite{Kato09}). $I_{\operatorname{diff}}$ is used to compute the total amount of work contributed by the carry-in parts, that is at most $I_{\operatorname{carry-in}}$ in Equation (\ref{eq:test}).  We do not detail it here since it will not be used in the remainder \footnote{Interested readers can refer to Section (4.3) in~\cite{Kato09}. $I_{\operatorname{carry-in}}$ is bounded by solving a knapsack problem in a greedy manner to fulfill as much as possible the interference rectangle by carry-in jobs.}. 

	Notice that $h_k$ is fixed while $w_k$ is not (since $\Delta_k$ is not determined). Thus, the previous condition must be checked for all values of $\Delta_k$. As a consequence, in order to use the test (Theorem~\ref{th:test}), $\Delta_k$ must be bounded to define a finite time interval to test possible values for $\Delta_k$ (Theorem~\ref{th:interval}).

	\begin{Theo}\label{th:interval}~\cite{Kato09}
		If Condition (\ref{eq:test}) is to be violated for any $\Delta_k$, then it is violated for some $\Delta_k \geq D_k$ satisfying Condition (\ref{eq:test}), where $C_{\operatorname{carry-in}}$ denotes $\sum_{\tau_i \in \tau_{\operatorname{carry-in}}} C_i$.

		\begin{equation}\label{eq:bound}
			\Delta_k \leq \frac{h_kC_k-\sum_{\tau_i \in \tau}(D_i-T_i)U_i\times \min(v_i,h_k)+C_{\operatorname{carry-in}}}{h_k - \sum_{\tau_i \in \tau}U_i\times \min(v_i,h_k)}
		\end{equation}

		where $\tau_{\operatorname{carry-in}}$ is the set of tasks with a carry-in job.
	\end{Theo}

	\section{Correctness Issues}\label{sec:countex}

	\begin{figure}
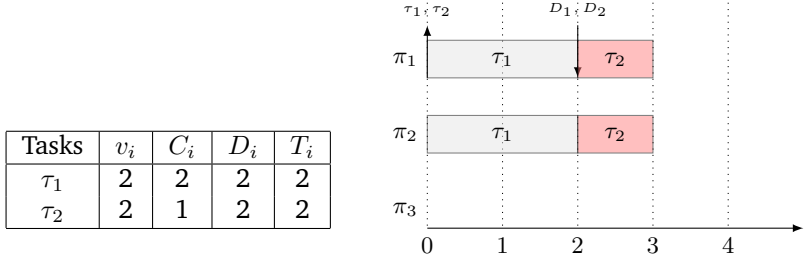

		\centering
		\begin{tabular}{c c }
			\subfigure{\raisebox{10mm}{
					\begin{tabular}{| c | c | c | c | c |}
						\hline
						Tasks & $v_i$ & $C_i$ & $D_i$ & $T_i$ \\ \hline
						$\tau_1$ & 2 & 2 & 2 & 2 \\
						$\tau_2$ & 2 & 1 & 2 & 2 \\\hline
					\end{tabular}
				}}&
				\subfigure{
					\begin{gantt}{4}{3}{1}{1}
						\activityrow{$\pi_1$}\request{$\tau_1,\tau_2$}\activity{$\tau_1$}{2} \deadline{$D_1,D_2$} \lateactivity{$\tau_2$}{1}
						\activityrow{$\pi_2$} \activity{$\tau_1$}{2} \lateactivity{$\tau_2$}{1}
						\activityrow{$\pi_3$} \lagtime{}{3}
					\end{gantt}
				}
			\end{tabular}
			\caption{Counter-Example: infeasible task set for 3-processor platforms.}
			\label{fig:fig}
		\end{figure}

		In this section, we show through a counterexample that problems arise when the schedulibility test presented in the previous section is applied to a counterexample task set. Then, we show that the necessary condition for a job to miss a deadline is not valid (i.e., Equation (\ref{eq:pb})). This will allow us to conclude that Theorems 1 and 2 that both exploit the contrapositive of Equation (\ref{eq:pb}) cannot define a valid sufficient schedulability test. We first present the counterexample task set.

		\paragraph{Counter-example definition.} Let us consider the feasible task set defined in Figure~\ref{fig:fig} and a platform with $m=3$ processors. Both tasks require simultaneously two processors and have a deadline of 2 units of time. According to Gang EDF, tasks $\tau_1$ and $\tau_2$ have equal priority since (i) the have the same relative deadline and (ii) they both require two processors \footnote{See Section 3 in~\cite{Kato09} for a complete presentation of Gang EDF.}. With no loss of generality, we assume that Gang EDF tie breaker gives the higher priority to $\tau_1$. Therefore, task $\tau_2$ will miss its deadline. Clearly, this task set is infeasible upon a 3-processor platform.

\subsection{Feasibility Interval}

		\paragraph{Applying the test on the counterexample.}	We analyze the task $\tau_2$. $\tau_2$ releases the problem job that misses its deadline at time 2 as depicted in Figure~\ref{fig:fig}. The feasibility interval is delimited by: $t_0=t_a=0$ and $t_d=D_2=2$; $\Delta_2=t_d-t_0=2$ and $A_2=\Delta_2-D_2=0$. The scenario $\Delta_2=D_2$ is the first scheduling point considered in the feasibility interval when applying Theorem~\ref{th:test}. The interference rectangle is: $w_2 = \Delta_2-C_2=2-1=1$ and $h_2 = m - v_2+1 = 3-2+1=2$.

		The first step in order to apply \textbf{[KAT]} test is to define the feasibility interval defined in Theorem~2. Next, we will only use the fact that $C_{\operatorname{carry-in}}\geq 0$\footnote{The inexistence of carry-in jobs for the considered task set will be proved in the remainder of this report.}. In Theorem~\ref{th:interval}, the numerator has always a positive value since $D_i \leq T_i$, $1\leq i \leq n$, and all used values are positive or zero. We will see that the denominator \emph{is not} positive in the counterexample. As a consequence, the schedulability test has to be applied over a time interval which has surprisingly a negative length.

	 The task under analysis is $\tau_2$, thus we set $k=2$ in the remainder. We need to bound $\Delta_2$ using Theorem~\ref{th:interval}. We prove hereafter that such a bound is negative for the counterexample. Consider the denominator of Inequality (\ref{eq:bound}): $h_k - \sum_{\tau_i \in \tau}U_i\times \min(v_i,h_k)$. As shown previously, we have $h_2=2$; this implies that $ \min(v_i,h_2) =2, 1\leq i \leq n$ and thus $h_2-\sum_{\tau_i \in \tau}U_i\times2=2-(\frac{2}{2}+\frac{1}{2})\times 2=-1$. As a consequence the  denominator is negative. Thus, the upper bound computed by Theorem~\ref{th:interval} of the time interval while checking the schedulability of a task has a negative length.

		Since, $\Delta_k <0$ implies that there is no $\Delta_k$ such that $\Delta_k \geq D_k$, then the sufficient schedulability test defined by Theorem~\ref{th:test} cannot be applied. According to our interpretation, such a situation cannot be interpreted as Theorem~\ref{th:test} \textit{is respected by default} but raises instead a correctness issue.

\paragraph{Insight.}

\begin{figure}[ht!]
			\begin{center}
				\includegraphics[width=10cm]{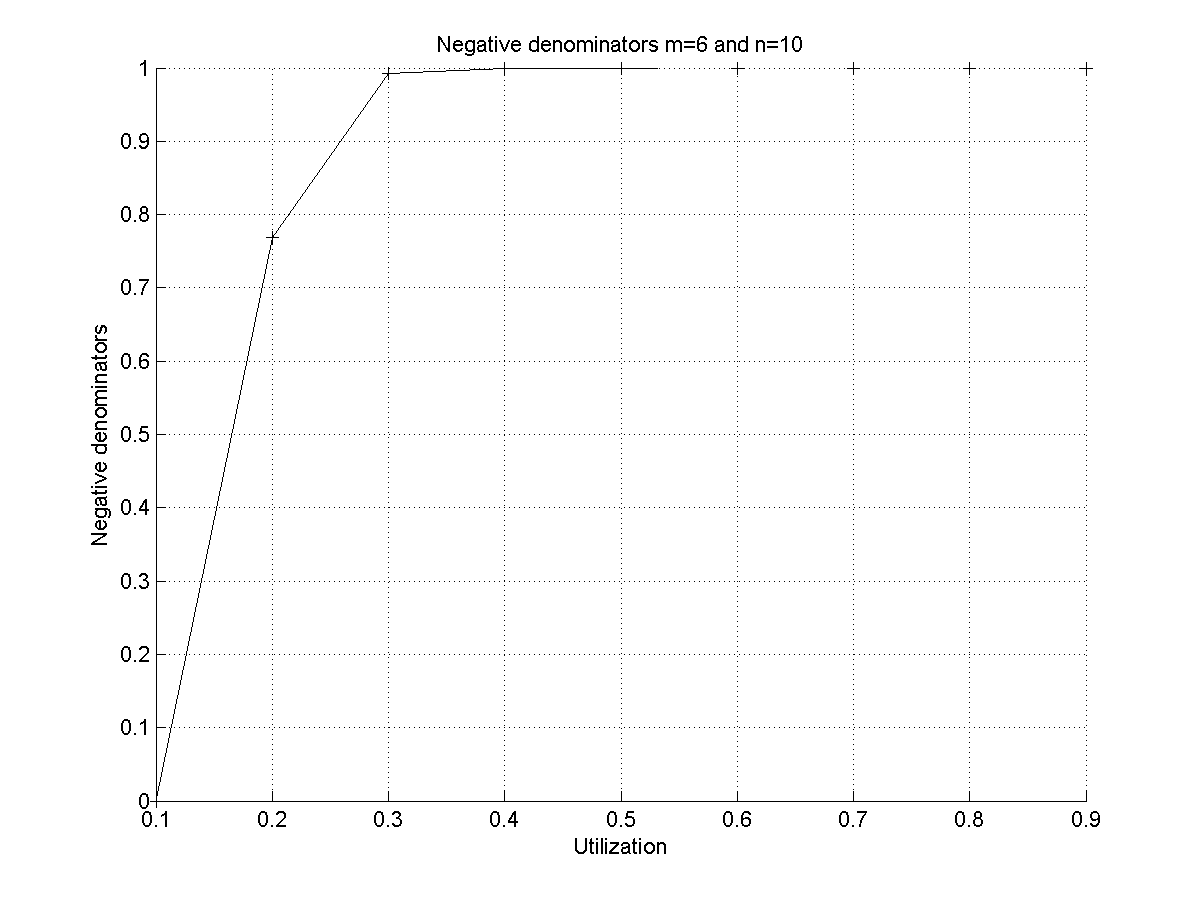}
				\caption{Task sets for which [KAT] cannot be applied}
				\label{fig:example1}
			\end{center}
		\end{figure}

		In~\cite{Kato09}, the last mathematical derivations performed to prove Theorem~\ref{th:interval} is incorrect. Precisely, if $(h_k - \sum_{\tau_i \in \tau} U_i \ times min(v_i, h_k)) \geq 0$ then:
		  \begin{equation}
		  \Delta_k \leq \frac{h_k C_k - \sum_{\tau_i \in \tau} \{  (D_i - T_i) \times \min(v_i, h_k) \} +C_{carry-in} }{ h_k - \sum_{\tau_i \in \tau} U_i \times min(v_i, h_k)}
		  \end{equation}
		  -   Otherwise:
		  \begin{equation}
		  \Delta_k > \frac{h_k C_k - \sum_{\tau_i \in \tau} \{(D_i - T_i) \times \min(v_i, h_k)\} +C_{carry-in} }{ h_k - \sum_{\tau_i \in \tau} U_i \times min(v_i, h_k)}
		  \end{equation}
		  Unfortunately most of the time we are in the second case and \textbf{[KAT]} cannot be applied. To clearly illustrate that purpose, Figure~\ref{fig:example1} depicts the number of task sets for which \textbf{[KAT]} cannot be used\footnote{Synthetic task sets with 10 tasks and 6 processors have been generated using the Stafford's algorithm.}. 
		  
\subsection{Necessary schedulability condition is incorrect}

		\paragraph{Applying the test on the counterexample.} 	In order to show that the necessary condition defined in Equation~(\ref{eq:pb}) for a task to miss its deadline is incorrect, we consider once again the task $\tau_2$ in the counterexample. 	First, we compute the interference bounds for the tasks without carry-in jobs:
		\begin{eqnarray*}
			I_1(\tau_1,\Delta_2)&=&\min\left(\left(\left\lfloor \frac{\Delta_2-D_1}{T_1} \right\rfloor +1\right)\times C_1,w_2\right) \times \min(v_1,h_2)\\
			&=&\min\left(\left(\left\lfloor \frac{2-2}{2} \right\rfloor +1\right)\times 2,1\right) \times \min(2,2) = 1 \times 2=2\\
			I_1(\tau_2,\Delta_2)&=&\min\left(\left(\left\lfloor \frac{\Delta_2-D_2}{T_2} \right\rfloor+1\right)\times C_2  -C_2,A_2\right) \times \min(v_2,h_2) \\
			&=&\min\left(\left(\left\lfloor \frac{2-2}{2} \right\rfloor+1\right)\times 1  -1,0\right) \times \min(2,2) = 0 \times 2=0\\
		\end{eqnarray*}

		Since both tasks have identical periods and deadlines, there is no carry-in job (i.e., jobs of $\tau_1$ and $\tau_2$ are released and have deadlines within the interval $[t_0,t_d)$). To prove this, let us show that $I_{\operatorname{diff}}(\tau_i)=0$, $1 \leq i \leq 2$. Let us compute $I_2(\tau_i)$:
		\begin{eqnarray*}
			I_2(\tau_1,\Delta_2)&=&\min\left(\left\lfloor \frac{\Delta_2}{T_1} \right\rfloor \times C_1+\min(C_1, \Delta_2 \mod T_1),w_2\right) \times \min(v_1,h_2)\\
			&=&\min\left(\left\lfloor \frac{2}{2} \right\rfloor \times 2+\min(2, 2 \mod 2),1\right) \times \min(2,2) = 1 \times 2=2\\
			I_2(\tau_2,\Delta_2)&=&\min\left(\left\lfloor \frac{\Delta_2}{T_2} \right\rfloor\times C_2 +\min(C_1, \Delta_2 \mod T_1)  -C_2,A_2\right) \times \min(v_2,h_2) \\
			&=&\min\left(\left\lfloor \frac{2}{2} \right\rfloor\times 1 + \min(1, 2 \mod 2) -1,0\right) \times \min(2,2) = 0 \times 2=0\\
		\end{eqnarray*}

		Hence there is no carry-in job since $I_{\operatorname{diff}}(\tau_i)=I_2(\tau_i)-I_1(\tau_i)=0$, $1 \leq i \leq 2$.

		Since there is no carry-in job (i.e., $I_{\operatorname{carry-in}}=0$), the total cumulative interference is $\sum_{\tau_i \in \tau} I_1(\tau_i) = 2$ and thus consequently the necessary condition defined in Equation (\ref{eq:pb}) is not satisfied since:
			\begin{eqnarray*}
				\sum_{\tau_i \in \tau} I_1(\tau_i) = 2 = w_2\times h_2
			\end{eqnarray*}
		To summarize, there is a deadline miss but the necessary condition defined in Equation (\ref{eq:pb}) is not satisfied. As a consequence, we conclude that Equation (\ref{eq:pb}) is not a valid necessary condition for a task to miss its deadline. Hence, Theorems~\ref{th:test} and \ref{th:interval} are not valid since they both use the necessary schedulability condition defined in Equation~\ref{eq:pb} as an initial claim.
		
		\paragraph{Insight.} This problem is in fact inherited from the [BAR] test. In a footnote in~\cite{Ber09,Ber11}, a technical problem has been exhibited and a simple solution is proposed to fix it. The detection of the error in \textbf{[BAR]} (\cite{Bar07}) was concurrent to the publication of \textbf{[KAT]}\footnote{Notice that in~\cite{Bar15} the erroneous version of [BAR] is presented.}. 
		
		The bound of the interference is not correct but can be easily fixed by:
		\begin{itemize}
			\item adding an $\epsilon$ in the interference bounds in $I_1$ and $I_2$ (resp. Equations (3) and (5) in~\cite{Bar07}),
			\item  or changing the inequality (8) in~\cite{Bar07} into a strict inequality.
		\end{itemize}
		
		Hence, using a strict inequality in Equation (\ref{eq:test}) (in Theorem~\ref{th:test}) will fix the problem that has been exhibited the previous counter-example while using [KAT] .
		
		\section{Discussion} \label{sec:discuss}

		This short note raises a major correctness issue in the schedulability test presented in~\cite{Kato09}. The first problem comes from the feasibility interval that may have negative length, even for infeasible task sets. Through numerical experiments, we shown that \textbf{[KAT]} cannot be applied. The second problem is inherited from the \textbf{[BAR]} that uses incorrect interference bounds. This latter problem can be corrected in \textbf{[KAT]} following the same principles used for correcting the similar problem in \textbf{[BAR]}.

		It is worth noticing that the detected problem in the necessary condition does not exist if the interference rectangle reaches its maximum height (i.e it is equal to $m$). This case only arises if every gang tasks uses exactly one processor at a time.  In this precise case, \textbf{[KAT]} schedulability test is valid and equivalent to \textbf{[BAR]}~\cite{Bar07} as explained in~\cite{Kato09}. 

		\bibliographystyle{acm}

		\bibliography{biblio}

	\end{document}